\documentclass[aip,cha]{revtex4-1}
\usepackage[T1]{fontenc}
\usepackage[latin9]{inputenc}
\usepackage{textcomp}
\usepackage{amsmath}
\usepackage{graphicx}

\makeatletter

\pdfpageheight\paperheight
\pdfpagewidth\paperwidth

\newcommand{\lyxmathsym}[1]{\ifmmode\begingroup\def\b@ld{bold}
  \text{\ifx\math@version\b@ld\bfseries\fi#1}\endgroup\else#1\fi}

\@ifundefined{textcolor}{}
{%
 \definecolor{BLACK}{gray}{0}
 \definecolor{WHITE}{gray}{1}
 \definecolor{RED}{rgb}{1,0,0}
 \definecolor{GREEN}{rgb}{0,1,0}
 \definecolor{BLUE}{rgb}{0,0,1}
 \definecolor{CYAN}{cmyk}{1,0,0,0}
 \definecolor{MAGENTA}{cmyk}{0,1,0,0}
 \definecolor{YELLOW}{cmyk}{0,0,1,0}
}

\makeatother


\begin{document}

\title{Chaotic motion of charged particles in toroidal magnetic configurations}

\author{Benjamin Cambon, Xavier Leoncini, Michel Vittot}

\affiliation{Aix Marseille Universit\'e, Universit\'e de Toulon, CNRS, CPT UMR 7332, 13288 Marseille, France }

\author{R\'emi Dumont, Xavier Garbet}

\affiliation{CEA, IRFM, F-13108 Saint-Paul-lez-Durance, France}
\begin{abstract}
We study the motion of a charged particle in a tokamak magnetic fi{}eld
and discuss its chaotic nature. Contrary to most of recent studies,
we do not make any assumption on any constant of the motion and solve
numerically the cyclotron gyration using Hamiltonian formalism. We
take advantage of a symplectic integrator allowing us to make long-time
simulations. First considering an idealized magnetic configuration,
we add a non generic perturbation corresponding to a magnetic ripple,
breaking one of the invariant of the motion. Chaotic motion is then
observed and opens questions about the link between chaos of magnetic
field lines and chaos of particle trajectories. Second, we return
to an axisymmetric configuration and tune the safety
factor (magnetic configuration) in order to recover chaotic motion.
In this last setting with two constants of the motion, the presence
of chaos implies that no third global constant exists, we highlight
this fact by looking at variations of the first order of the magnetic
moment in this chaotic setting. We are facing a mixed phase space
with both regular and chaotic regions and point out the difficulties
in performing a global reduction such as gyrokinetics.
\end{abstract}


\maketitle

\begin{quotation}
Contrary to old studies, most of recent research dealing with hot magnetized fusion
plasmas rely on numerical simulations. In the special case of a tokamak magnetic
configuration, the majority of the numerical codes are based on the gyrokinetic theory.
One of the assumptions made by this theory is to consider the magnetic moment of the
particles moving in the magnetic field as an exact invariant of the motion. The straight
consequence is to consider that all particle trajectories are integrable for axisymmetric
configuration and no electric field. This assumption enables to make global reduction of
the phase space and allows for faster numerical simulations. In fact, the magnetic
moment is often an adiabatic invariant and it can present variations over very large-time
scales. These remarks lay the ground for possible presence of Hamiltonian chaos in
particle trajectory and a non-constant magnetic moment.
In this paper, we solve numerically the motion of charged particles including the cyclotron
gyration using Hamiltonian formalism in the sixth dimensional phase space, without
using any assumption. We take advantage of a symplectic integrator allowing us to make
long-time simulations. First, considering an axisymmetric magnetic configuration, we
add a non-generic perturbation corresponding to a specific magnetic ripple, breaking one
of the invariant of the motion. Chaotic motion is then observed whereas magnetic field
lines are still integrable. So, we underline that the link between the two notions is not
automatic. Second, we observe chaos of particle trajectories even in an axisymmetric
configuration of the magnetic field. For this purpose, we study the limit case in which the
major radius of the tokamak is infinite. The geometry becomes cylindrical. We tune the
winding profile of magnetic field lines in order to create a separatrix in the effective
Hamiltonian of the cylindrical integrable system. We then perturb the system by adding
some curvature and returning back to the toroidal configuration. The main result is that
the chaotic region of the phase space of the system grows when the major radius of the
''tokamak'' decreases. The presence of a chaotic trajectories and mixed phase-space
implies that the magnetic moment of the charged particles is not a global constant of
the motion. We show its variations and remark that, in the case of a chaotic motion, this
variation are significant and entice us to be careful before performing a global gyrokinetic reduction.

\end{quotation}

\section{Introduction}

The motion of charged particles in a magnetic field is a classical
issue of dynamical systems and plasma physics, as for instance the
fifty years old study made in\cite{Bogoliubov61,Kruskal62}, before
a time when numerical simulations became mainstream, or more recent
approaches as for instance \cite{CaryBrizard2009,NouvelElement1}.
Usually, the main properties of particle trajectories
strongly depend on the form of magnetic field lines, making global
features difficult to determine without using basic approximations.
That's why, the theory of plasma confinement into a tokamak magnetic
field is often studied using gyrokinetic theory. One of the assumptions
made by regular codes based on this approach is that the magnetic
moment $\lyxmathsym{\textmu}$ of confined particles is constant\cite{Porcelli01,eriksson,Egedal00}.
The straight consequence is to consider that particle trajectories
are integrable for axisymmetric configuration and
no electric field.

Indeed, trajectory of a charged particle in a tokamak idealized magnetic
field is a six dimensions system in which we can consider two exact
invariants, the energy and the angular momentum related to the toroidal
invariance of the field. Moreover, the magnetic moment is said to
be an adiabatic invariant. So, in good conditions, $\mu$ is a quasi-constant
on short time but has some variations over very large-time scales
(see for instance \cite{Neishtadt86,Neishtadt97,Benisti07,Leoncini09,Neishtadt2013}).
Due to Arnold-Liouville Theorem, integrability of charged particle
trajectory requires three exact commuting invariants \cite{Arnold}.
Suppose that this three quantities commute, which is not obvious,
we can transform the original system into an integrable approximation.
However this may change the nature of particles trajectories. The
link between chaotic motion and variations of the magnetic moment
is an important topic \cite{Drag76,Weitzner99} and these remarks
lay the ground for possible presence of Hamiltonian chaos of particle
trajectory and a non-constant magnetic moment even in an idealized
magnetic case. 

In this paper, we integrate the Hamilton equations through a symplectic
code to be able to solve on one part the cyclotron motion on short
time, and on an other part, to make long-time simulations, without
approximations. Moreover, contrary to classical approach based on
Runge-Kutta method, the main advantage of the symplectic codes is
to keep constant the phase space volume and
eventually the invariants of the motion over large-time scales, allowing
us to study adiabatic variation of the magnetic moment.

Using this property, the question we would like to answer is the presence
of Hamiltonian chaos in charged particle motion in a toroidal magnetic
field, studying two different cases. First, we consider
a non generic magnetic ripple modeling the finite
number of coils surrounding the tokamak chamber, and
breaking one of the invariant of the motion. In the
second case, we want to create chaos to particle trajectories keeping
an idealized magnetic field but looking for chaos in a six-dimensional
system is not simple. In order to find the chaotic regions of the
phase space, we decide to study the limit case in which the aspect
ration (major radius vs minor one) of the tokamak tends to infinity.
In this limit, the tokamak becomes a cylinder and the motions of the
charged particles are integrable. The method is to create a separatrix
in this simple system using the fact that the poloidal part of the
magnetic field in the tokamak chamber is not a well-known function.
Then, we will return to the toroidal case and look at the alteration
of the integrability of the particle trajectories around the separatrix.
In the same time, this second method highlights variations of the
magnetic moment.

\section{Model}

\subsection{Basic equations}

\begin{figure}
\centering{}\includegraphics[width=8cm]{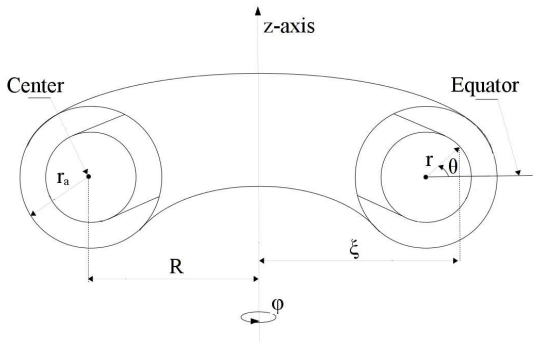}\caption{Toroidal geometry and notations\label{fig:Toric-geometry}}
\end{figure}

A schematic view of the toroidal geometry is presented on fig \ref{fig:Toric-geometry}.
 In the $(r,\:\theta,\:\varphi)$ coordinates, we shall first consider
the tokamak magnetic field of the following form \cite{Dif-pradelier11}:

\begin{equation}
\mathbf{B}=\frac{B_{0}R}{\xi}(\hat{\mathbf{e}}_{\varphi}+f(r)\:\hat{\mathbf{e}}_{\theta})\:,\label{eq:champ_B}
\end{equation}
in which $\xi=R+r\cos(\theta)$, $\hat{\mathbf{e}}_{\varphi}$
and $\hat{\mathbf{e}}_{\theta}$
are the unit vectors associated respectively to the $\varphi$ and
$\theta$ directions. The toroidal component along $\hat{\mathbf{e}}_{\varphi}$
of the magnetic field is due to external coils around the tokamak
chamber whereas the plasma generated current inside the tokamak leads
to the creation of the poloidal component $\hat{\mathbf{e}}_{\theta}$.
In this approximation the function $f(r)$ represents the amplitude
of the poloidal field as a function of the radius and it is directly
connected to the so-called safety factor $q$ of
principal importance in tokamak physics : 
\begin{equation}
f=r/q\xi\approx r/qR\:.\label{eq:F_enfonction deS}
\end{equation}
In Eq.~(\ref{eq:F_enfonction deS}) for simplicity we shall use the
approximation in order to simply have $q=q(r)$.
Usually, the amplitude of the poloidal component is estimated at about
$10\%$ of the global magnetic field but the safety factor is not
an a priori fixed function and can actually be tuned in some machines.
So, respecting the flux conservation of the magnetic field $\nabla\cdot\mathbf{B}=0$,
we have the freedom to choose the function $f(r)$. It is easy to
show that any smooth function depending only on $r$ satisfies the
flux conservation constraint.

Associated to Eq.~(\ref{eq:champ_B}), a vector potential can be
chosen respecting the Coulomb gauge :

\begin{equation}
\mathbf{A}(r)=B_{0}\frac{F(r)}{\xi}\:\hat{\mathbf{e}}_{\varphi}-B_{0}R\:\log(\frac{\xi}{R})\:\hat{\mathbf{e}}_{z}\:,\label{eq:Potentiel_A}
\end{equation}
where $F(r)=\int^{r}f$. This choice of
the magnetic field and associated potential vector corresponds to
nested circular magnetic surfaces, which is a good approximation in
the large aspect ratio limit. In a more realistic situation, we may
have for instance to take into account that the magnetic surfaces
have a slowly drifting center.

Given the magnetic field we consider the dynamics of a particle with
charge $e$ and mass $m$, in this magnetic field. The Hamiltonian
of the system writes

\begin{equation}
H=\frac{(\mathbf{p}-e\mathbf{A})^{2}}{2m}\:,\label{eq:Hamiltonien}
\end{equation}
where $\mathbf{p}=(p_{x},p_{y},p_{z})$
and $\mathbf{\mathbf{x}}=(x,y,z)$ form three pairs of
canonically conjugate variables. The associated equations of motion
are :
\begin{equation}
\begin{cases}
\dot{\mathbf{x}} & =(\mathbf{p}-e\mathbf{A})/m\:,\\
\dot{\mathbf{p}} & =\frac{e}{m}(\nabla\mathbf{A})\cdot(\mathbf{p}-e\mathbf{A})\:.
\end{cases}\label{eq:Equation of motion}
\end{equation}
 We readily notice in (\ref{eq:Equation of motion}) that the particle
velocity $\dot{\mathbf{x}}$ is given by $\mathbf{v}=(\mathbf{p}-e\mathbf{A})/m$.

\subsection{Chaos, invariants and Adiabatic approximation}

Looking for chaotic aspects of particle motions leads to look at the
invariants of this three-dimensional system. Indeed, finding three
independent invariants commuting with each other would imply integrability
in the sense of Arnold-Liouville \cite{Arnold}. The energy of the
particle is an exact invariant equal to $H=\frac{1}{2}m\mathbf{v}^{2}$.
Then in the case of the magnetic field (\ref{eq:champ_B}), the system
is invariant by rotation around the $\hat{\mathbf{e}}_{\varphi}$
axis. This entails that the conjugated angular momentum $M$ is a
constant of the motion (Noether Theorem). We can define it using the
Lagrangian 
\begin{equation}
\mathcal{L}=\frac{1}{2}m\mathbf{v}^{2}+e\:\mathbf{A}\cdot\mathbf{v}\:,\label{eq:Lagrangian_def}
\end{equation}
by

\begin{equation}
M=\frac{\partial\mathcal{L}}{\partial\frac{d\varphi}{dt}}\:.\label{eq:definition_of_M}
\end{equation}
Using this definition (\ref{eq:definition_of_M})
and the value of the vector potential in the proposed gauge, we obtain

\begin{equation}
M=\zeta\:\mathbf{p\cdot\hat{\mathbf{e}}_{\varphi}}\:.\label{eq:Def_simple de M}
\end{equation}
Besides these two exact constants of the
motion, we need a third integral in order to get integrable motion.
The third constant that is often used is the magnetic
moment $\mu$ defined at leading order by :

\begin{equation}
\mu=\frac{m\mathbf{v}_{\perp}^{2}}{2B}\:,\label{eq:Mu_0}
\end{equation}
where $\mathbf{v}_{\perp}$ denotes the
component of the velocity $\mathbf{v}$ perpendicular to the magnetic
field $\mathbf{B}$. The magnetic moment~(\ref{eq:Mu_0}) of a gyrating
particle is used as a constant even though it is often only an adiabatic
invariant \cite{Gardner1959,Boozer80,Cary86,Littlejohn81}. Assuming
that these three integrals are in involution we end up with an integrable
system. Besides the involution, it is not obvious that the adiabatic
invariant is actually a real invariant, in fact it is likely that
depending on the magnetic configuration, this assumptions breaks down
in some regions of phase space (see for instance
\cite{Drag76,Weitzner99}), leading to local Hamiltonian chaos. The
breaking of this assumption could be problematic, especially since
in the last fifteen years a considerable effort had been devoted to
gyrokinetics. Hot magnetized plasmas are usually well described using
a Maxwell-Vlasov or Vlasov-Poisson description. This implies working
with a time dependent particle density function in six-dimensions.
In gyrokinetics theory the adiabatic invariant is used to reduce the
dimensionality of phase space. This allows to not only to move to
an effective $4+1$ dimensional phase space, but allows also to avoid
resolving in time the fast gyration around field lines (cyclotron
frequency), both of these features are very appealing when aspiring
at a full fledged numerical simulation of magnetized fusion plasmas
in realistic conditions. In this paper we do not consider the consequences
of the presence of Hamiltonian chaos on this theory, we however will
look for the breaking of the adiabatic invariant. In this spirit we
consider so-called fast-particles, meaning in typical tokamak conditions,
we consider particles whose energy is from $10keV$ to around $3.5MeV$,
which is about the energy of $\alpha-$particles resulting from the
fusion of Deuterium and Tritium nuclei. Looking at the variation of
$\mu$ given by Eq.~(\ref{eq:Mu_0}) on time scales
shorter than the cyclotron period, we observe a non constant function,
which at first approximation looks like a sinusoid with
a specific period. Given this fact, in what follows we now consider
the average of $\mu$ on one or more cyclotron period as the effective
adiabatic invariant, but keep the same notation.

In order to look for chaos, a possibility would be to actually find
traces of adiabatic chaos, meaning chaos on adiabatic time scales.
So, looking for the main characteristics
of the motion of the particle, we have to study long-term variations
of $\mu$. The requirements for a numerical approach are thus that
we must resolve short time variation meaning the fast cyclotron motion
but we also have to simulate on very long time to be able to capture
possible fluctuations of the magnetic moment. We therefore need to
use a symplectic approach.

\subsection{Numerical scheme}

It is well known that using a classical Runge-Kutta scheme leads to
important long-time errors. Indeed the scheme is usually not symplectic,
hence the time steps does not preserve efficiently phase-space volume,
in this case long-time studies can lead to the emergence of sinks,
sources and attractors, which do not belong to the realm of Hamiltonian
chaos see for instance field lines in \cite{Leoncini06}. In order
to avoid this difficulty, numerical simulations have been performed
using the sixth order Gauss-Legendre symplectic scheme discussed in
\cite{McLachlan92}. Since we needed pairs of canonically conjugated
variables, we remained in Cartesian coordinates. This
symplectic construction ensures the conservation of the symplectic
volume and so, the Hamiltonian nature of the purpose. Moreover, we
observe on Fig.~\ref{fig:Toric-geometry} that this
construction ensures also the stability of the energy
and the angular momentum of the particle even for very large times
and chaotic trajectories (see Sec.~\ref{sec:Impact-of-the})
\begin{figure}
\centering{}\includegraphics[width=8cm]{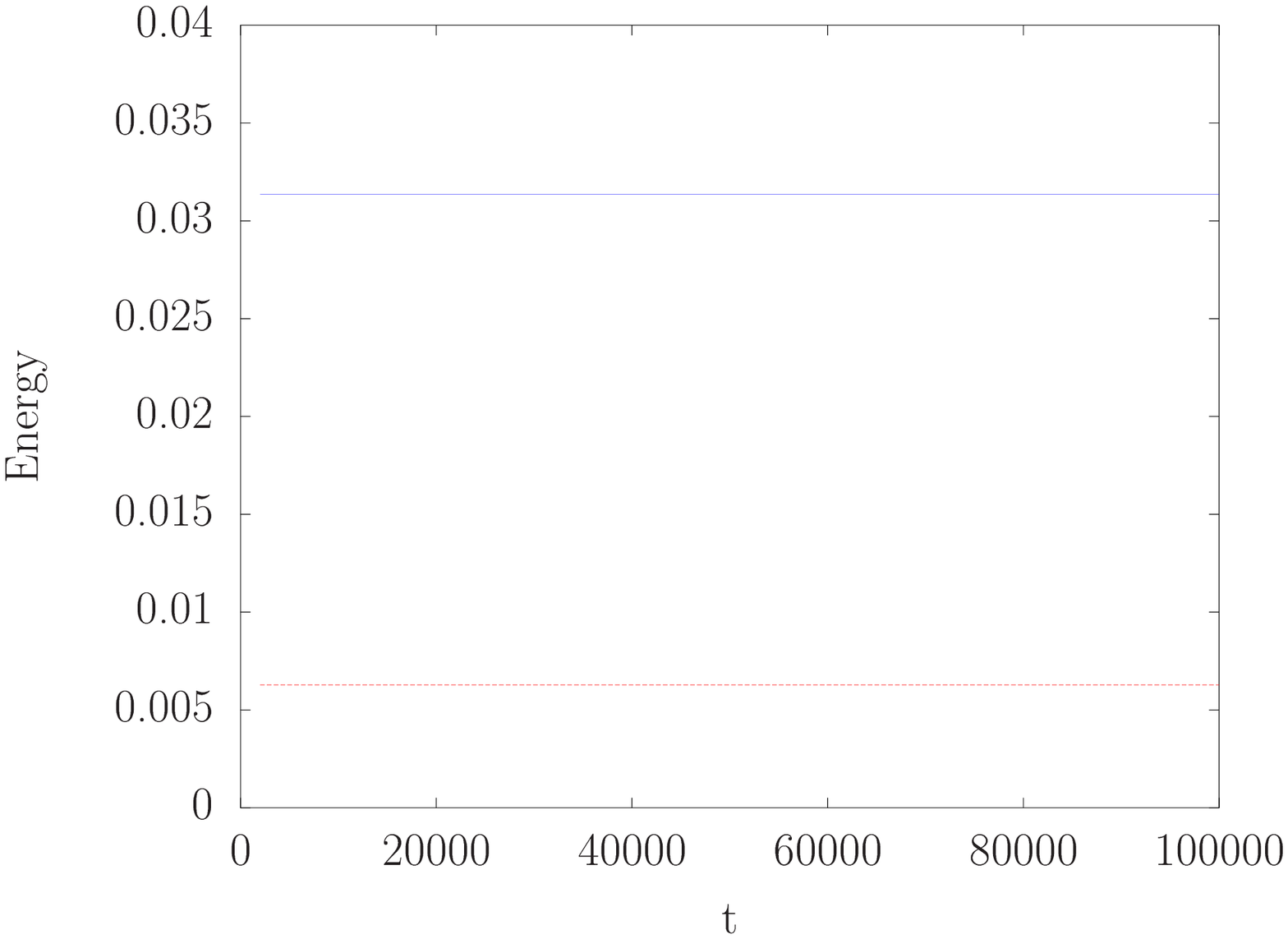}\\
\includegraphics[width=8cm]{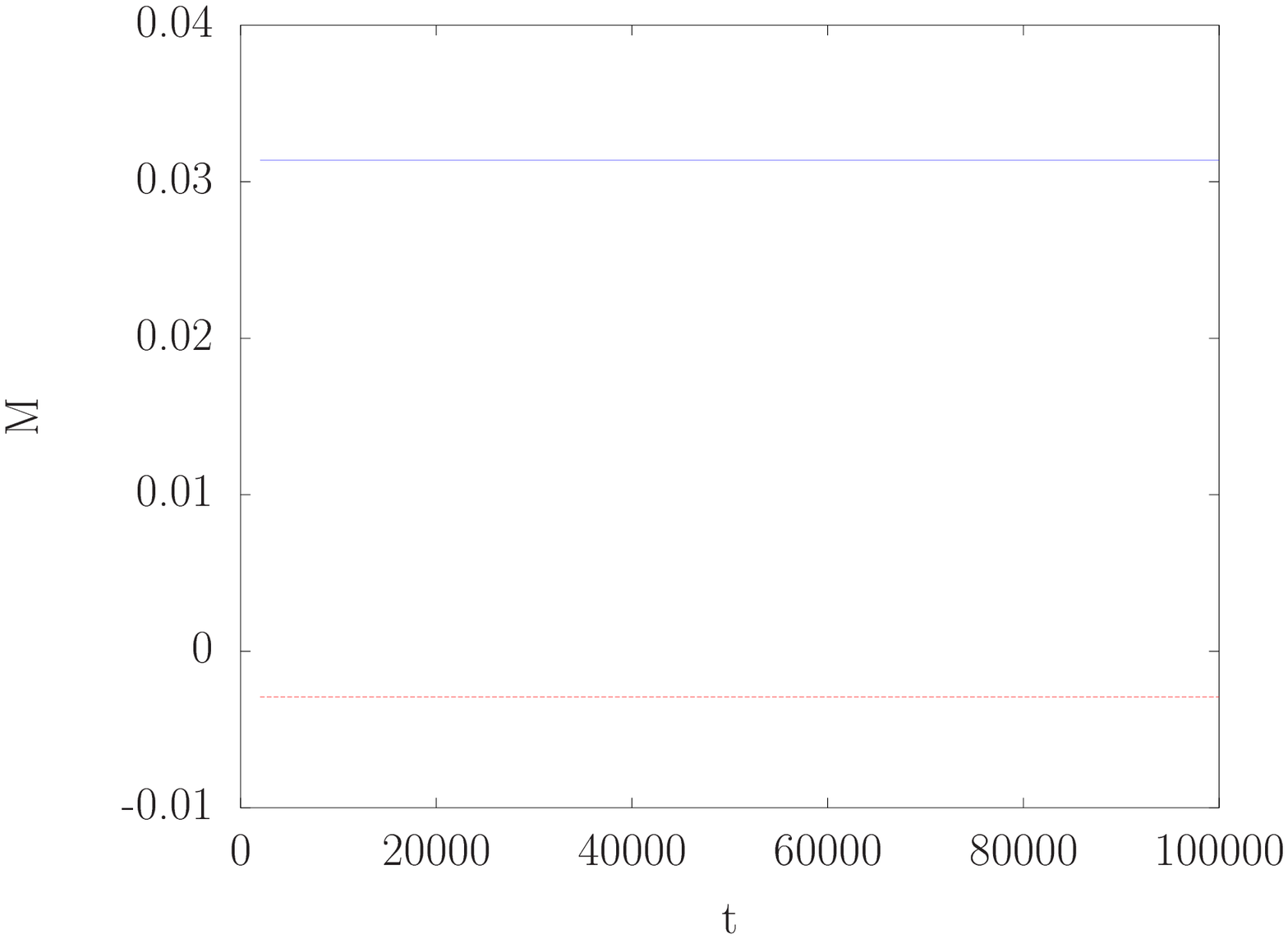}\caption{top: Stability of the energy of the particule in two
different cases : one integrable trajectory choosen in the case of
$f=0$ (full line) and another picked up from a chaotic trajectory
(dashed line) in the case $R=3$ (see Sec.~\ref{sec:Impact-of-the})
Bottom : Same comparison with the angular momentum M }
\end{figure}
 Last but not least, in order to minimize numerical errors due truncations
we use dimensionless variables. This allows to
use variables whose variations remain in a finite range. Our choice
is performed as follows, noting $\widetilde{r}$, $\widetilde{t}$
and $\widetilde{B}$ our new variables. We write :

\begin{eqnarray}
\widetilde{r} & = & r/r_{a}\:,\label{eq:Adimensionnement}\\
\widetilde{t} & = & \omega_{c}t=\frac{m}{eB_{0}}t\:,\\
\widetilde{\mathbf{B}} & = & \mathbf{B}/B_{0}\:,
\end{eqnarray}
where $r_{a}$ is the minor radius (see
Fig.~\ref{fig:Toric-geometry}), $\omega_{c}$ is the cyclotron frequency,
and $B_{0}$ is the typical magnetic intensity at the center of the
torus. Thereafter, we will only consider these dimensionless variables,
and for clarity and readability we drop the $\tilde{}$ and note $(r,\: t,\: B)$
the dimensionless variables.

\begin{figure}
\centering{}\includegraphics[width=8cm]{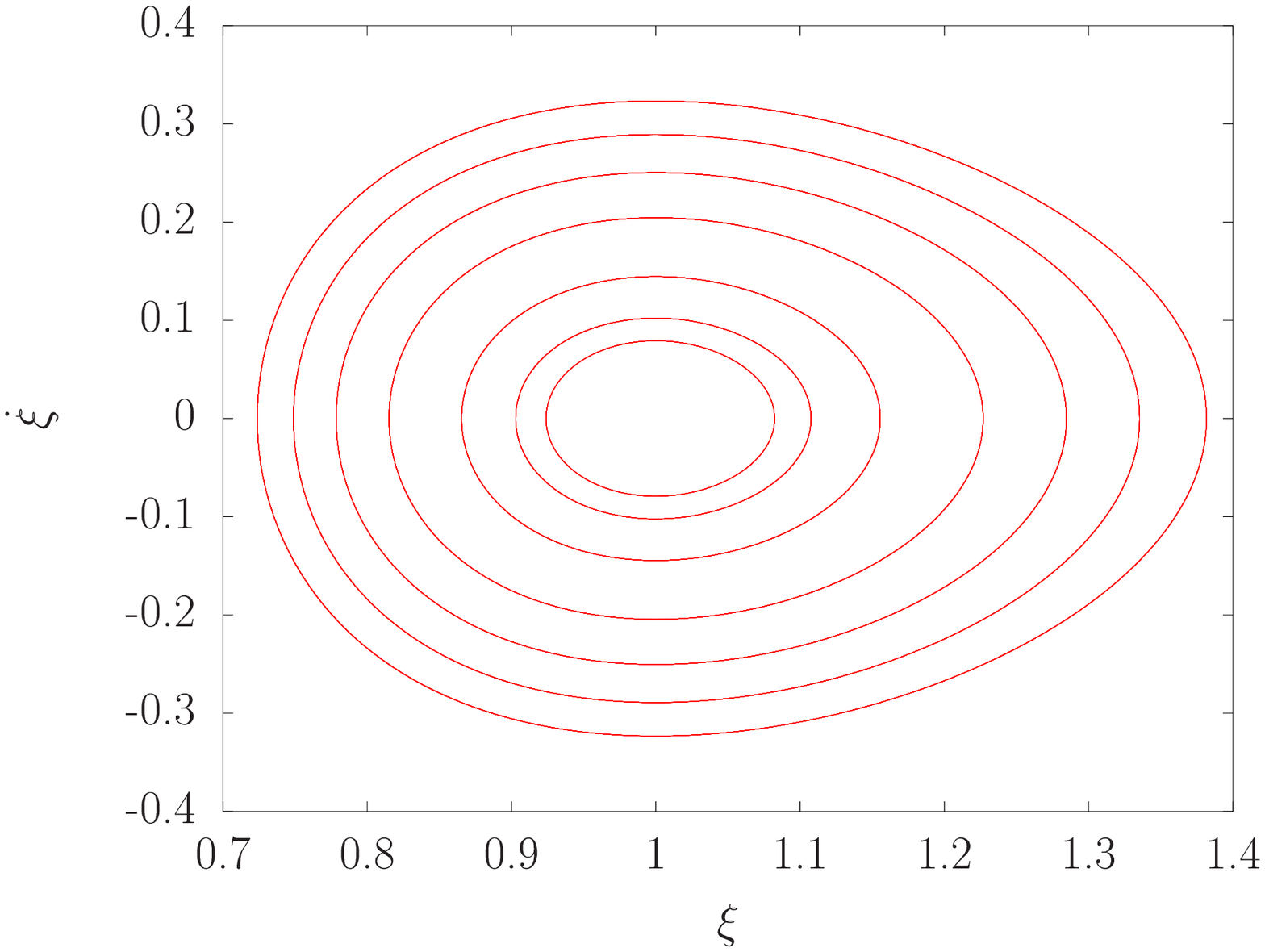}\\
\includegraphics[width=8cm]{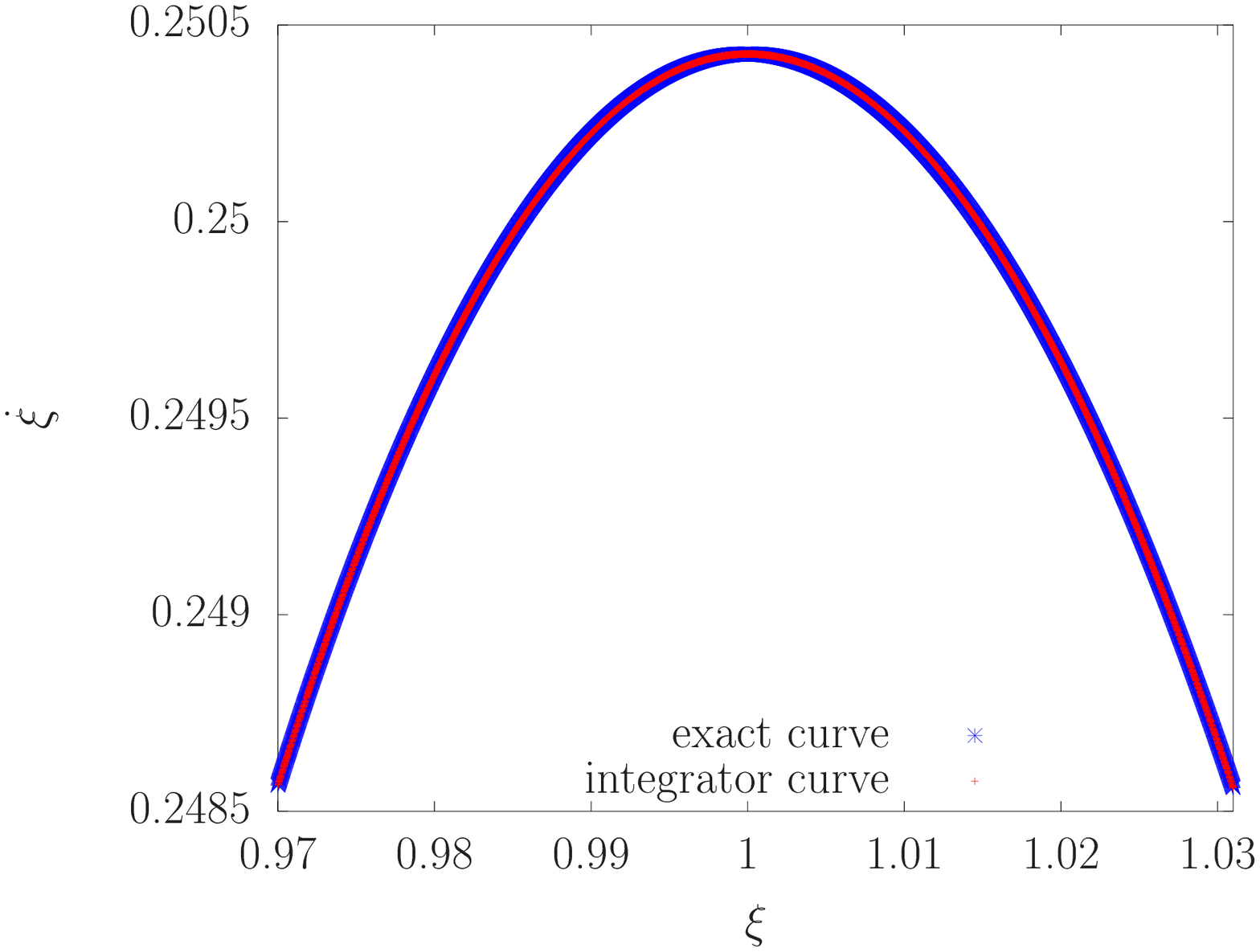}\caption{Top: phase portrait in $(\xi,\dot{\xi})$ coordinates for the simple
case without plasma in the tokamak chamber. Bottom: zoom and comparison
between a numerically computed trajectory and the analytical one.
Numerical integration is performed for $4.\:10^{7}$ time steps using
a $\delta t=0.01$. \label{fig:f=00003D0-1}}

\end{figure}

In order to validate the accuracy of our numerical scheme and test
our code, we first consider the particular case in
which there is no plasma inside the tokamak. In this configuration,
the poloidal magnetic field is negligible. This property entails that
the magnetic field depends only on $\xi$. So, in polar coordinates
$(\xi,\varphi,z)$, $\mathbf{B}=R\hat{\mathbf{e}}_{\varphi}/r_{a}\xi$.
The application of Newton's second law, which computation is detailed
in annex~\ref{Annex1}, demonstrates the drift of the particle in
$z$-direction (the direction depends on the sign of the charge).
In fact the system resumes to an integrable one and we end up with
an effective Hamiltonian :

\begin{equation}
H_{eff_{0}}=\frac{\dot{\xi^{2}}}{2}+\frac{C^{2}}{2\xi^{2}}+\frac{ln^{2}(\xi)}{4}+C'ln(\xi)\:,\label{eq:Hamilton_pure_torus}
\end{equation}
in which $C=\xi^{2}\dot{\varphi}$ and
$C'$ are constants, which depend on the particle's initial conditions.
A comparison between the $(\xi,\dot{\xi})$ phase portrait obtained
by our numerical scheme and the one obtained by resolving the quadrature,
using the Hamiltonian~(\ref{eq:Hamilton_pure_torus}) is shown in
Fig.~\ref{fig:f=00003D0-1}. 
 We can notice an excellent agreement between
the two curves, which validates our code. To be more specific, the
step time is chosen to be $\delta t=0.01$, which means we should
have about $100$ points to resolve the gyration motion. The records
in Fig.~\ref{fig:f=00003D0-1} cover a period of $4.\:10^{7}$ time
steps with $R=3$, $C=1$ and $C'=0$. Further analysis of the Hamiltonian~(\ref{eq:Hamilton_pure_torus}),
leads as conclusions that there are no separatrix in phase space and
that the equilibrium point corresponds to a linear drift of the particle
while all other trajectories are helical. As can be noted in Fig.~\ref{fig:f=00003D0-1},
when we zoom on one curve, we can compare in blue the exact trajectory
and in red the motion computed by the integrator. The two methods
give the same results with a precision better than our recorded digits,
meaning better than $10^{-5}$. Another check of the numerical accuracy
of our computation will be performed in Sect.~\ref{sec:Impact-of-the}.

\section{Chaotic particle trajectories using ripple effect\label{sec:Chaotic-trajectories-using}}

\subsection{The plasma with cylindrical symmetry first part}

\begin{figure}

\begin{centering}
\includegraphics[width=8cm]{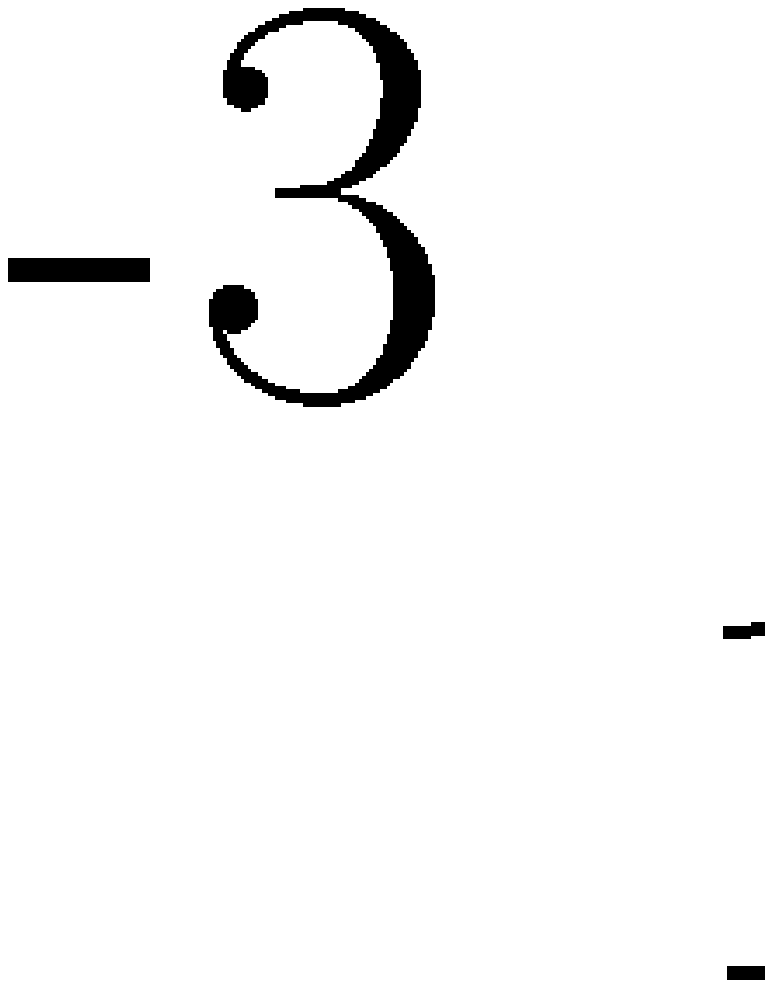}
\par\end{centering}

\caption{$(\xi,z)$ section obtained with $d\mu/dt=0$ with no ripple. Four
different trajectories with different energies but same speed directions
are displayed. Motion appears as integrable, we notice the presence
of a separatrix, distinguishing to types of orbits: so called passing
orbits and banana orbits. \label{fig:classic_xi_z}}

\end{figure}

In the more general case when a plasma is present in the tokamak,
the $(\xi,\dot{\xi)}$ plane looses its specificity because the magnetic
field $\mathbf{B}$ also depends on $r$ through the expression of
the safety factor. In order to choose a configuration (\ref{eq:f_realiste}),
we take a function that satisfies $q(0)=1$,
that presents a maxima and that respects $\lim_{r\rightarrow\infty}q(r)=0$~:

\begin{equation}
q(r)=\frac{1+ar}{1+r^{2}}\:,\label{eq:f_realiste}
\end{equation}
we recover the magnetic profile and the
function $f(r)$ using the approximate expression in (\ref{eq:F_enfonction deS}).
In the following, we will fix the constant $a=10$. It becomes then
challenging to visualize trajectories in the sixth-dimensional phase
space. Indeed, the existence of two exact first integrals leaves room
for possible chaos, moreover the expected cyclotronic motion of charged
particles is likely to blur sections. Choosing a good section for
visualizing trajectories becomes therefore challenging. The first
problem is related to the choice of the initial conditions in a six-dimensional
phase space. If we focus on fast particles corresponding to the ashes
of fusion, their initial conditions are not explicitly known, however
we can consider that their energy (mostly a kinetic one) is more or
less known and about $3.5\text{MeV}$ and that their initial location
is close to the center of the chamber, where the plasma is dense and
hot and fusion reactions take place. However, the direction of the
speed is uncertain, leaving us with to two independent parameters,
the norm being fixed by a chosen energy. Then we have to find a suitable
section in which we can distinguish integrable and chaotic trajectories.
If we only consider a classical section such as
the poloidal $(\xi,z)$ plane launching particles
with a fixed angle, things are not clear due to the Larmor gyration,
we end up with ``thick'' trajectories for all
types of initial conditions. In order to better visualize we have
considered two possible options. In a first attempt
we computed the time average of $\mu(t)$ over a large number gyrations
\begin{equation}
\bar{\mu}(\tau)=\frac{1}{\tau}\int_{0}^{\tau}\mu(t')dt'\:.\label{eq:average mu}
\end{equation}
Typically we computed $\bar{\mu}$ during a given initial time of
the trajectory. We then performed a section $(\xi,z)$ and recorded
positions each time $\mu(t)=\bar{\mu}$ and $d\mu/dt>0$. We then
get clear thin and distinct integrable trajectory. However for some
initial conditions and a different type of magnetic
field, we noticed a slow drift in $\mu(\tau)$ such that we could
not record any data for large portions of the trajectory as $\mu(t)$
was not crossing the initially computed finite time
average$\bar{\mu}$. In order to circumvent this last problem and
noticing that $\mu(t)$ always had a fast oscillating component, we
finally settled for the following section: we only record points when
reaching local maxima of $\mu(t)$, meaning we plot the points when
$d\mu/dt=0$ and $d^{2}\mu/dt^{2}<0$. As a consequence, plotted points
correspond to one Larmor gyration. The resulting$(\xi,z)$ section
is presented in Fig.~\ref{fig:classic_xi_z}, where different trajectories
are represented and have different values of the energy but the same
initial speed direction.
 On this this section, we see two
different types of trajectories: the elliptic ones (corresponding
to so-called passing particles) and the bananas, separated by a separatrix.
They all appear to be integrable, and looking for Hamiltonian chaos
is not easy. Indeed in this setting, because the localization of the
separatrix appears to depend on the initial speed direction, we can
only represent particles with the same speed directions and location
initial conditions on a given section, else we actually end up with
a projection and characterized by apparent crossing
of trajectories. In order to look for Hamiltonian chaos one of the
solutions is to add a perturbation into the magnetic field.

\subsection{The ripple effect}

In order to trigger chaos more efficiently, we have introduced so
called-ripple effect \cite{Yushmanov90}. This effect is mainly due
to the finite number of coils around the tokamak and leads to the
breaking of the toroidal invariance of the magnetic field, with for
instance a dependence of the magnetic field on
the toroidal angle $\varphi$. A direct consequence of the symmetry
breaking is that the angular momentum $M$ ceases to be an invariant,
the accessible phase space becomes larger and chaos can occur even
if another ``hidden'' constant of the motion related to the magnetic
moment exists.

\begin{figure}

\begin{centering}
\includegraphics[width=8cm]{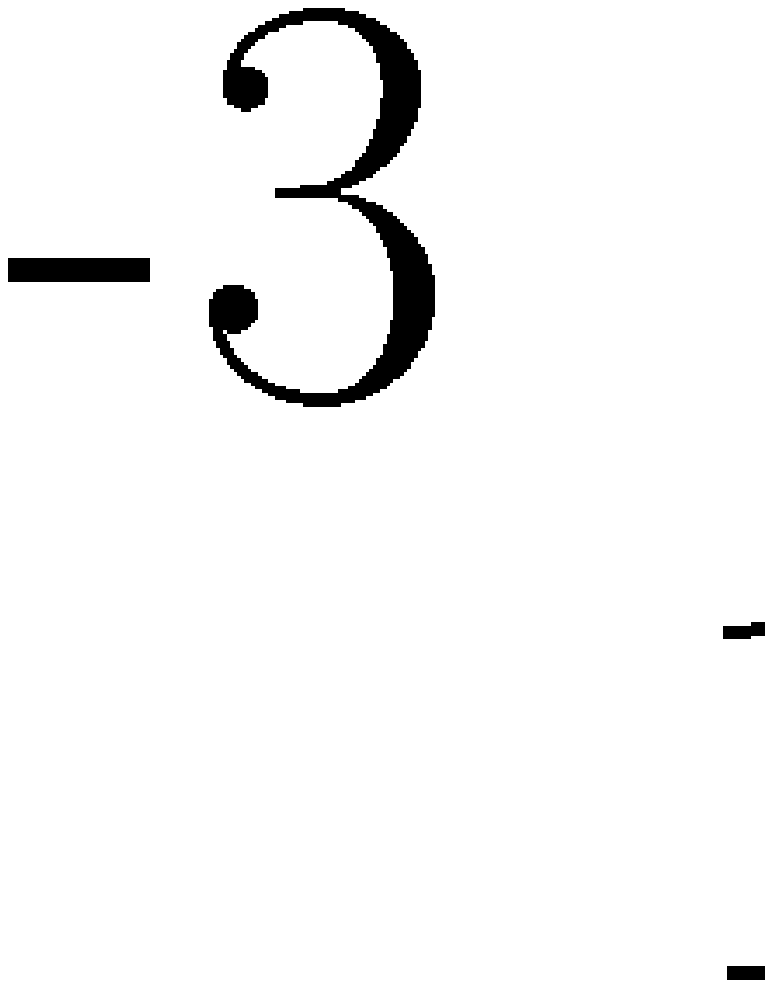}\\
\includegraphics[width=8cm]{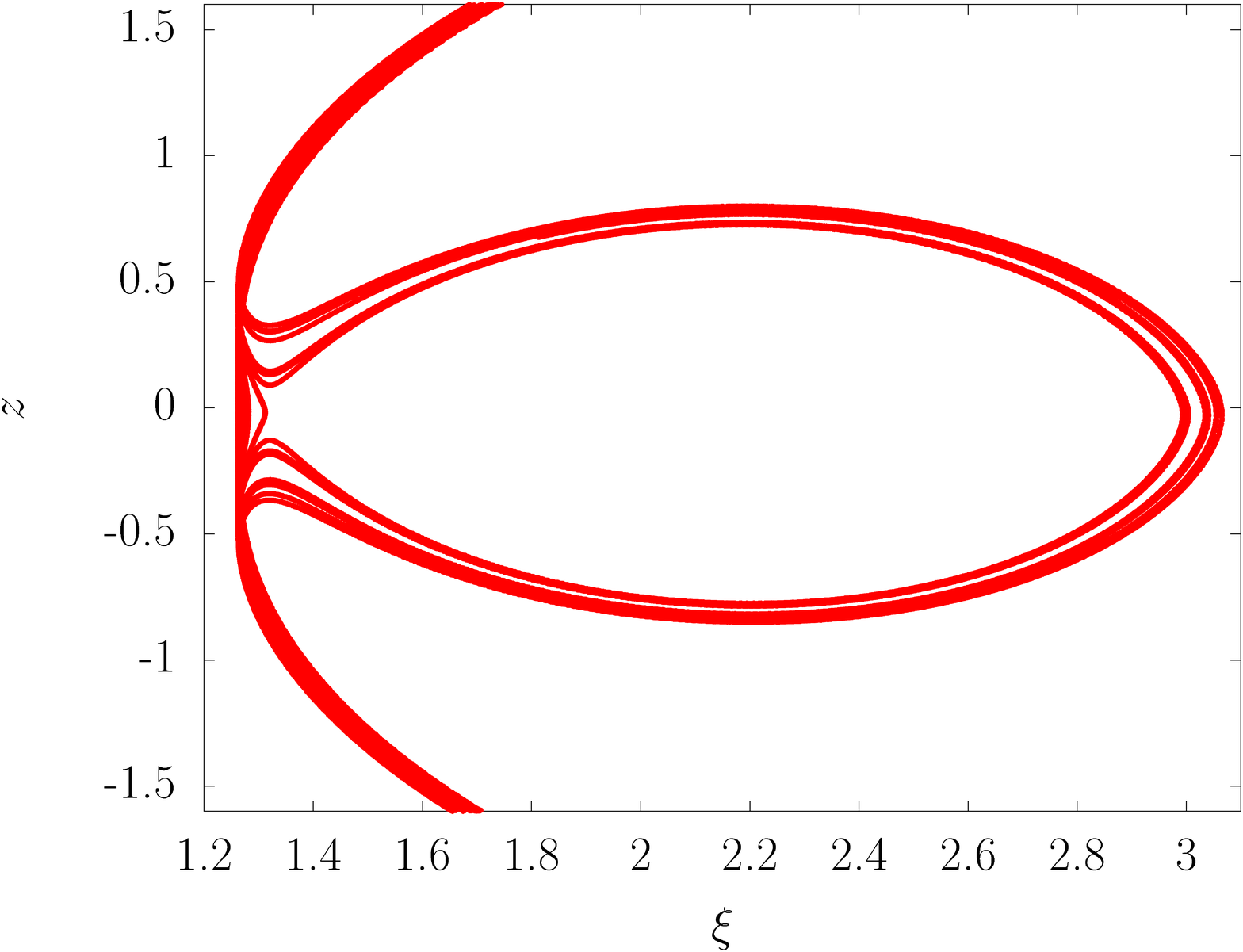}
\par\end{centering}

\caption{Top: $(\xi,z)$ section obtained with $d\mu/dt=0$, and one trajectory
with ripple and $\varepsilon=0.2$. Bottom: zoom of the section near
the separatrix, we see the jumps between the passing orbits and the
banana ones.\label{fig:chaotic-trajectory-ripple}}

\end{figure}

In order to remain simple we shall consider a very specific and actually
non generic perturbation. This translate into a vector potential modified
as: 

\begin{equation}
\mathbf{A}(r)=\frac{F(r)}{\xi}(1+\delta\cos(k\varphi))\hat{\mathbf{e}}_{\varphi}-\log(\xi)\hat{\mathbf{e}}_{z}\:,\label{eq:ripple_potential_vector}
\end{equation}
in which $\delta$ represents the amplitude
of the perturbation entailed by the ripple and $k$ the number of
coils. In the case of ITER, $k=18$ and$\delta$ is estimated to $10^{-3}$
(the expression is of the perturbation is though different). As mentioned
the breaking of one of the exact invariant of the motion makes easier
the search of chaotic motion because the trajectory is now confined
in a five dimensional phase space. We show in Fig.~\ref{fig:chaotic-trajectory-ripple}
the motion of one charged particle in the perturbed magnetic field,
but for a perturbation amplitude $\delta=0.2$. 
Contrary to the idealized magnetic case,
chaos shows up and is characterized by separatrix breaking. Starting
from a banana motion, the particle trajectory is able to cross the
separatrix to become an elliptic passing one and vice versa. We though
would like to insist that a priori we remain in a situation with mixed
phase space. In the sense that if we start at a sufficiently low energy,
far from the separatrix with an elliptic trajectory, the motion does
not present sign of chaos at least for the length of time during which
we computed and observed such trajectories.

\subsection{Link between chaos of magnetic field lines and chaotic trajectories
of charged particles}

One of the usual cartoons that is used when describing magnetic confinement,
is that particles have a fast gyration around magnetic field lines,
and if the Larmor radius is small, particles follow more or less field
lines. When looking at the previous sections in Fig.~\ref{fig:classic_xi_z},
we directly see that this is likely not the case. However this remark,
lead to numerous studies on the chaoticity of magnetic field lines,
having in mind a picture very much like the advection of passive tracers
in a stationary three dimensional flows, where passive tracers simply
follows velocity field lines \cite{morrisson2000,Firpo2013,Evans2010,Babiano94,Basu03,Benkadda97,Beyer2001,Carreras03,Ciraolo04,Leoncini06,Portela2007}.
The tokamak being a three dimensional object, it has been shown for
a long time that field lines of flux conservative fields are generically
chaotic in three dimensions. Assuming particles follow field lines
it becomes natural to infer that studying the chaos of field lines
would have a direct impact on particles confinement and trajectories.
In the ideal tokamak, the magnetic field lines are not chaotic, due
to the existence of the invariance by rotation of an angle $\varphi$.
In the chosen magnetic fields, the lines winds helically around a
given torus. Breaking the invariance, would lead to a real three dimensional
system and chaos of field lines. It is in this sense
that the perturbation chosen in Eq.~(\ref{eq:ripple_potential_vector})
is non generic. Indeed in this last expression the magnetic field
still does not depend on $r$, nor has a component along $\hat{\mathbf{r}}$,
hence the magnetic field lines still gently wind around a torus, the
variation of intensity is compensated by a non uniform winding, which
allows to keep a constant flux across a given disk of radius $r$.
Field lines are therefore still integrable, and a Poincaré section
of these lines would not differ from the perfectly invariant case. 

We insist therefore on the fact that despite the regularity of the
magnetic field lines, we end up with chaotic particle trajectories.
The connection between the two being therefore not obvious. Given
this remark, we shall reconsider the case with an rotational invariance,
and see if we can also find chaotic motion of charged particle. In
order to illustrate this we reconsider our point of view and start
with a cylindrical plasma tube instead of an empty toroidal solenoid.

\section{Impact of the toroidal geometry on particle trajectories\label{sec:Impact-of-the}}

\subsection{Infinite torus}

In this section, we use another approach to create
chaotic trajectories in a toroidally invariant
magnetic field. For this purpose we start with
a cylindrical magnetic geometry, which is the limit, when $R$ tends
to infinity, of the toroidal system. In this approximation, $\hat{\mathbf{e}}_{\varphi}$
becomes a stationary vector and the idealized magnetic field can be
rewritten as:
\begin{equation}
\mathbf{B}=B_{0}(\hat{\mathbf{e}}_{\varphi}+f(r)\hat{\mathbf{e}}_{\theta})\:.\label{eq:B_cylindrique}
\end{equation}
A calculation described in appendix~\ref{Annex2} leads to and integrable
system with a new effective Hamiltonian:

\begin{equation}
H_{eff}=\frac{\dot{r}^{2}}{2}+\frac{C''^{2}}{2r^{2}}+\frac{r^{2}}{8}+\frac{F^{2}(r)}{2}=E_{c}+V_{0}(r)+\frac{F^{2}(r)}{2}\:,\label{eq:Effective Hamiltonian_cylindr}
\end{equation}
in which 
\begin{equation}
C''=r(v_{\theta}+r/2)\label{eq:Expression of Cprimeprime}
\end{equation}
is a constant. The potential $V_{0}(r)$ tends to infinity when $r$
tends to zero or infinity. Between these limits, it admits a minimum
at $r_{0}=\sqrt{2C''}$. We can thus tune the function $F(r)$ in
order to generate a separatrix in this effective
Hamiltonian corresponding to the cylindrical system. For this purpose
we create a local maximum of the effective potential
energy. We remark that choosing a $F(r)$, corresponds to fixing a
specific so called $q-$profile of the tokamak plasma
magnetic confinement. We then perturb the system
by returning to a toroidal geometry, decreasing the value of $R$
to a finite value. A possible choice for $F$ can
be :

\begin{equation}
F=ar^{2}\exp(-\frac{r\text{\texttwosuperior}}{c\text{\texttwosuperior}})\:,\label{eq:F_separatrix}
\end{equation}
with $a=30$ and $c=\sqrt{100}$. The details
of the initial conditions which are fixed by $C''$ and $H_{eff}$
are explained in appendix~\ref{Annex2}. Given
these choices, the figure \ref{fig:cylindric-geometry} shows the
phase space $(r,\dot{r})$ of the effective system for a proton with
the particular condition $C''=\sqrt{10^{-5}}$. 
\begin{figure}

\centering{}\includegraphics[width=8cm]{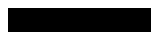}\\
\includegraphics[width=8cm]{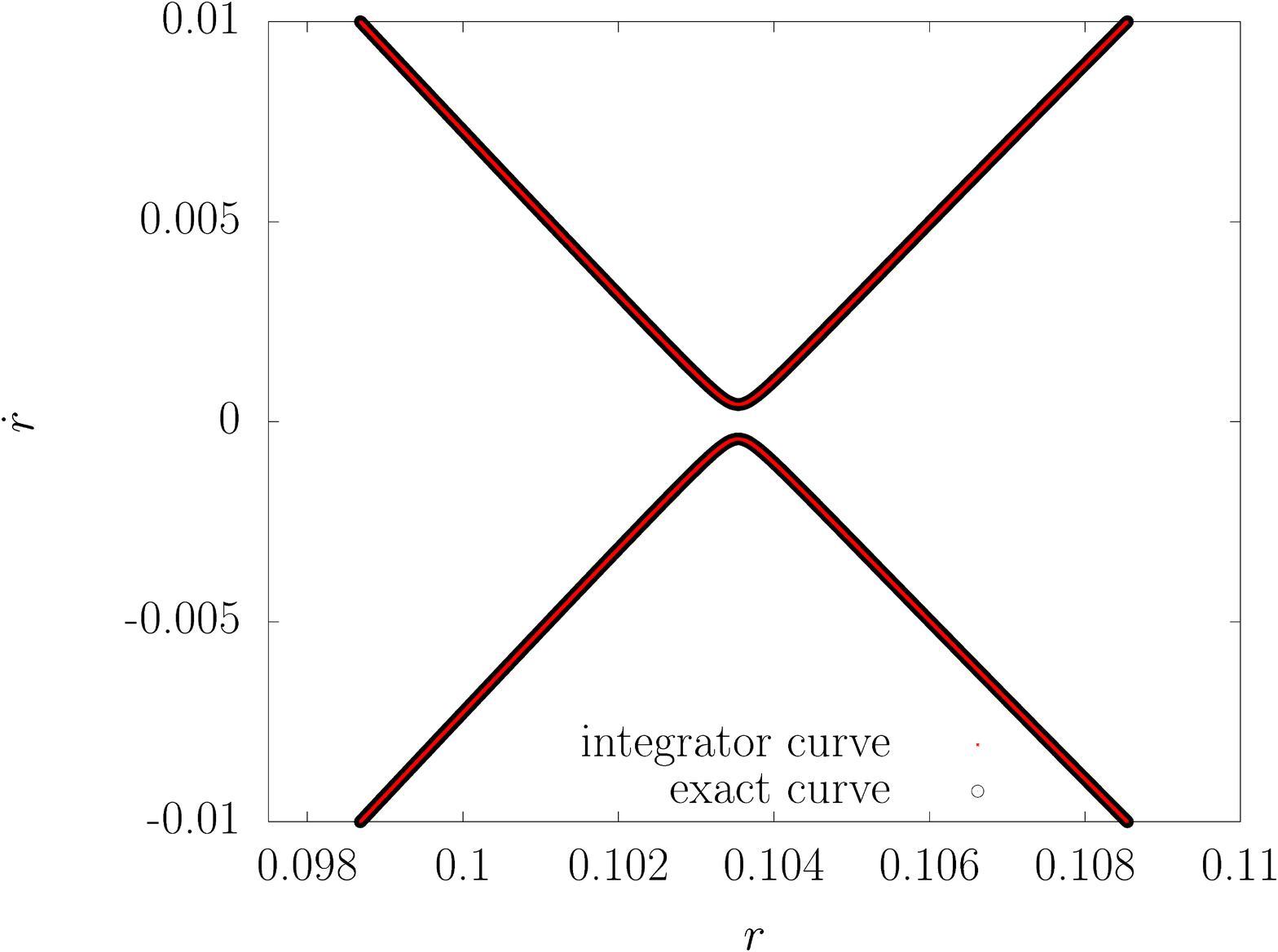}\caption{Top: phase space for a particle in the cylindric magnetic field governed
by Hamiltonian~(\ref{eq:Effective Hamiltonian_cylindr}). Bottom:
zoom near the separatrix and comparison between analytical curve solved
by quadrature and numerical integration.\label{fig:cylindric-geometry}}

\end{figure}
As expected, the motion depicted on Fig.~\ref{fig:cylindric-geometry}
shows that the particle trajectory in this type of magnetic field
is integrable. We visualize the separatrix between two areas corresponding
to the potential wells and the rest of the phase space for particles
with an energy of about $E=600keV$. This gives us another framework
to successfully test the accuracy of our numerical integration, especially
in the vicinity of the separatrix as described in Fig.~\ref{fig:cylindric-geometry}.

Looking for chaos, we shall now perturb this system
going back to the toroidal geometry and study the impact of the curvature
onto trajectory properties.

\subsection{Particle trajectories in toroidal geometry }

As mentioned we consider the toroidal aspect with finite large radius
of the tokamak as a perturbation of the cylindrical system. We now
consider finite radius $R$, and will slowly increase the ``perturbation''
by decreasing the value of $R$. We are eager to compare the motion
of a charged particle in the cylindrical geometry to the one in the
toroidal system. For this purpose we shall visualize the particle
trajectories in the $(r,\dot{r})$ plane. In order to perform a clean
section, considering the action-angle variables associated to the
integrable Hamiltonian~(\ref{eq:Effective Hamiltonian_cylindr}),
section are usually best performed at constant action. Indeed if we
only consider a projection on the $(r,\dot{r})$ plane, as previously,
the Larmor radius causes that we end up with thick trajectories for
all trajectories. Since computing the action is not a straightforward
task, we shall use the fact that for integrable systems $H=H(I)$,
a section at constant $I$, corresponds to a section at constant energy.
For this purpose, we reconsider the effective Hamiltonian $H_{eff}$~(\ref{eq:Effective Hamiltonian_cylindr}),
which in the cylindrical case can be rewritten as
\begin{equation}
H_{eff}=\frac{v_{r}^{2}}{2}+\frac{C''^{2}}{2r^{2}}+\frac{r^{2}}{8}+\frac{v_{\varphi}^{2}}{2}\:,\label{eq:Heff-2}
\end{equation}
and given the expression of $C''$~(\ref{eq:Expression of Cprimeprime})
we notice it corresponds to the actual kinetic energy of the system
up to some constant ($2C''$). However for a finite radius this Hamiltonian
presents temporal variation and becomes a time dependent function
$H_{eff}(t)$. Its expression does not change but, in the toroidal
system, $C''^{2}$ is not a constant anymore. So performing a section
at constant action corresponds to perform a section at constant $H_{eff}$,
which corresponds to a section taken at constant $C''$, since of
course the kinetic energy of a particle is a constant of the motion.
\begin{figure}
\centering{}\includegraphics[width=8cm]{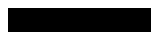}\\
\includegraphics[width=8cm]{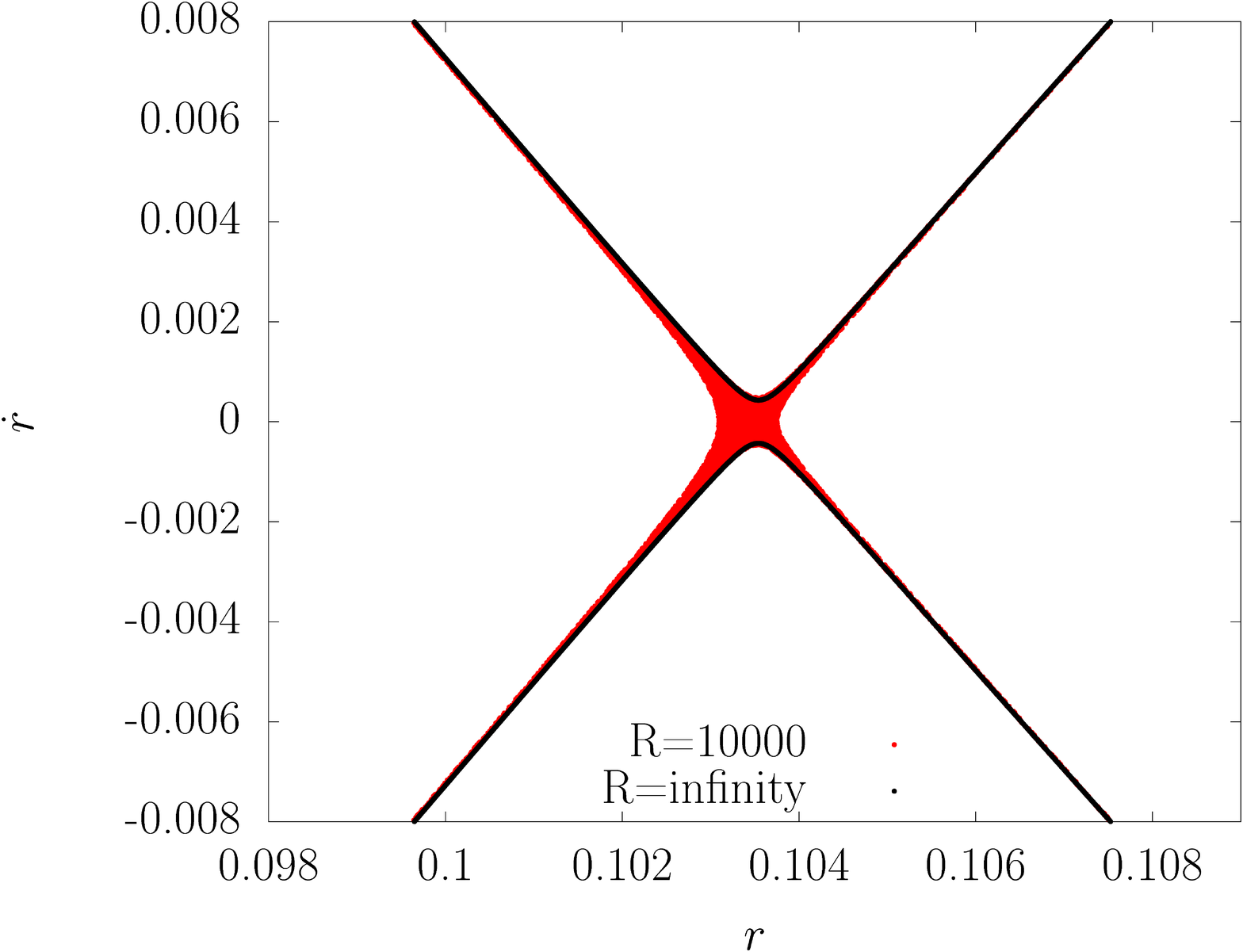}\caption{Top: $(r,\dot{r})$ section for $H_{eff}=Constant$, $R=10000$, $E=600keV$.
Initial conditions are identical to the ones taken in the cylindrical
case near the separatrix in Fig.~\ref{fig:cylindric-geometry}. Bottom:
zoom of the trajectory near the unstable fix point, the integrable
trajectory is drawn for comparison. Hamiltonian separatrix chaos emerges.
\label{fig:R10000_entier}}
\end{figure}

We first start with a small perturbation, namely $R=10000$ (in our
dimensionless units $R$ is measured in minor radii). Results are
displayed in Fig.~\ref{fig:R10000_entier}. In fact most of the trajectories
remain regular, but when we consider a particle with initial conditions
close to separatrix in the integrable cylindrical case, represented
in Fig.~\ref{fig:R10000_entier} something happens.
In fact, the overall picture has the same aspect as the cylindrical
phase space (Fig.~\ref{fig:cylindric-geometry}), except if we zoom
around the unstable equilibrium point, there the
dynamics becomes very different and we see the emergence of a chaotic
region, often dubbed a stochastic layer. As in Sect.~\ref{sec:Chaotic-trajectories-using},
the chaotic region is created by the breaking of the separatrix, in
the current case we remain however with an axisymmetric magnetic
field. As such the presence of chaos in the charged
particle trajectory indicates that the system is non-integrable and
de facto rules out the existence of a third constant of the motion,
for instance one related to the magnetic moment. 
\begin{figure}
\centering{}\includegraphics[width=8cm]{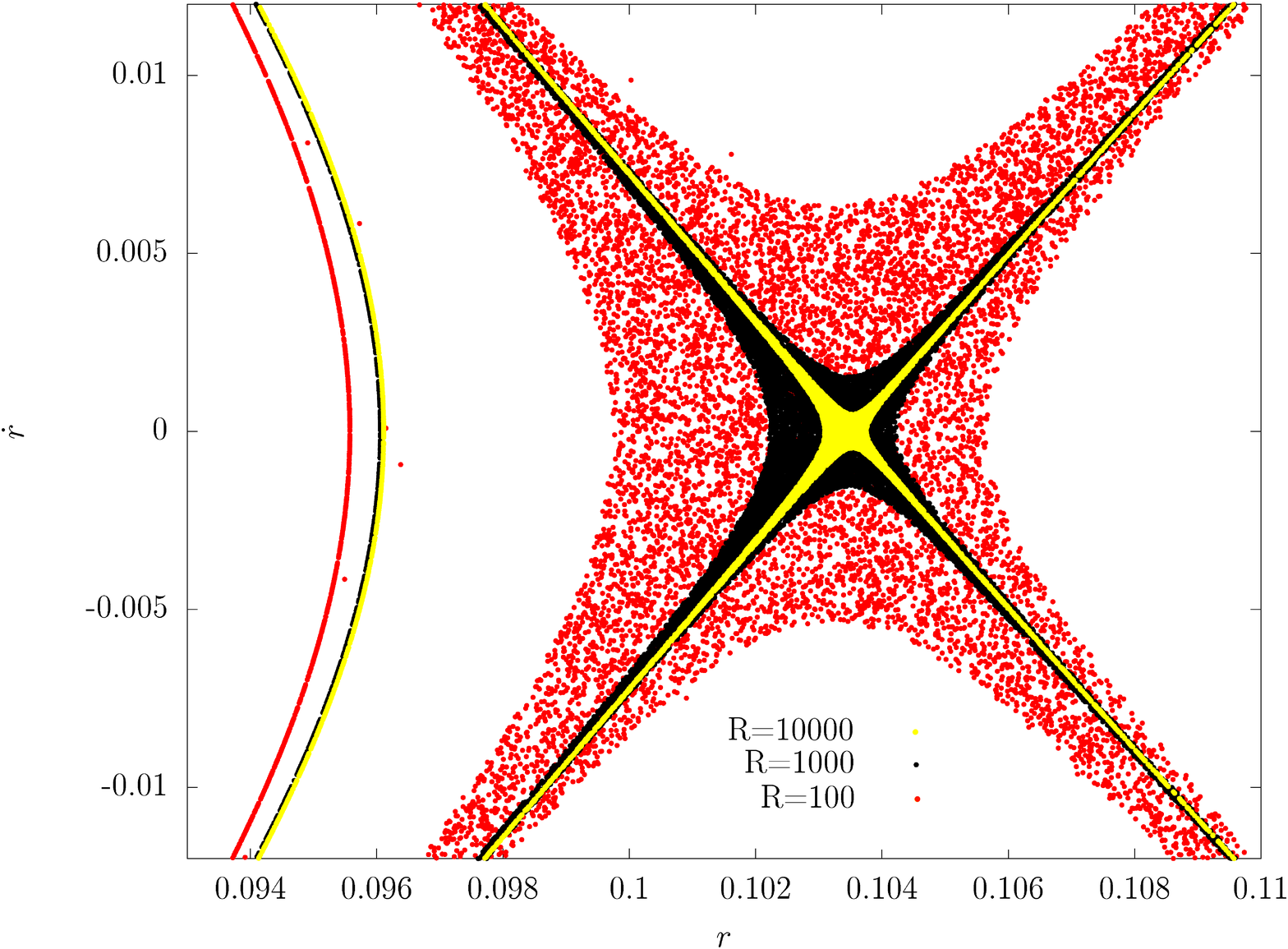}\caption{$(r,\dot{r})$ sections for $H_{eff}=Constant$ and $R=10000,\,1000,\,100$
. We see an increase of the stochastic layer as the major radius is
reduced. For $R=100$, we start to see some problems with the section
(see Fig.~\ref{fig:R100-1}). For comparison tori corresponding to
identical initial conditions and the different radii are drawn in
the regular region. Final time is $t=6\,10^{4}$. For each regular
trajectory the initial conditions are: $x_{0}=R+0.05$, $y=0$, $z=0$,
$vy=-F(r(t=0))$, $vz=-(C''/(x-R)-(x-R)/2.0)$, $E=590\text{keV}$,
$v=\sqrt{2E/m}$, $vx=\sqrt{v\text{\texttwosuperior}-vy\text{\texttwosuperior}-vz\text{\texttwosuperior}}$
, for the chaotic trajectories they are the same but with $E=600\text{keV}$.
\label{fig:R100}}
\end{figure}

\begin{figure}
\centering{}\includegraphics[width=8cm]{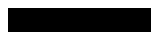}\\
\includegraphics[width=8cm]{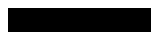}\caption{Top: $(r,\dot{r})$ section for $H_{eff}=Constant$, with $R=10$,
$E=600keV$. Initial conditions are identical to the previous cases,
meaning the one taken in the cylindrical case near the separatrix
in Fig.~\ref{fig:cylindric-geometry}. Bottom, same section but with
$R=3$. We notice in both plots what looks like an increase of chaos,
but that also that the sections at constant action appears as non
satisfactory anymore. \label{fig:R100-1}}
\end{figure}

In order to have a clearer picture we increase the perturbation and
we decrease progressively $R$ from $R=10000$ to $R=100$. The global
picture is displayed in Fig.~\ref{fig:R100}.
We consider the same initial condition and display the section of
one trajectory, we confirm the expected trend that increasing the
perturbation, corresponding to a reduction of the aspect ration $R$
leads to a growth of the chaotic region in the phase space. In order
to be more conclusive, we as well display regular trajectories obtained
in these three settings using the same initial condition as well.
The emergence of chaos is a priori not a numerical artifact. When
further decreasing $R$, we notice that an other phenomenon appears
when we pass from $R=100$ to $R=10$; as can been
seen on the section for $R=10$ and $R=3$ depicted in Fig.~\ref{fig:R100-1}
and a global apparent increase of chaos in the charged
particle trajectory emerges. It is though likely
that in this range, the section at constant action is not suitable
to correctly describe the system, this is relatively surprising as
the perturbation is still relatively small and we are actually reaching
the aspect ration of real machines such as the one of ITER for instance.
The problem of creating a good section in this region needs still
to be solved, and one of the problems of remaining with section made
using the cylindrical action could be related to monodromy issues.

\begin{figure}
\centering{}\includegraphics[width=8cm]{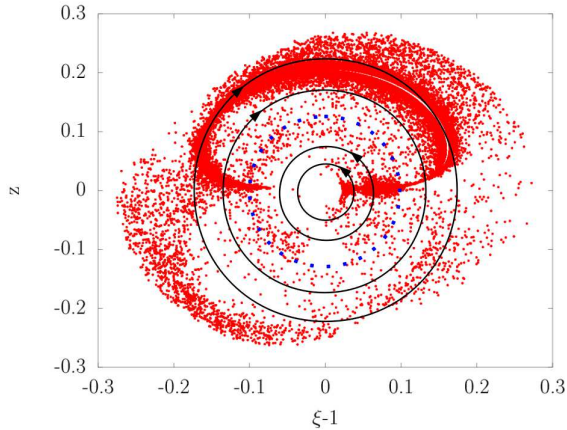}\\
\caption{Poloidal section of magnetic field lines. The field is given by Eq.~(\ref{eq:champ_B})
and $f$ is obtained from Eq.~(\ref{eq:F_separatrix}). Field lines are purely toroidal and thus periodic  at  $r=c=0.1$, where  the winding of the magnetic field lines changes orientation. We insist
on the fact that this does not depend on the value of $R$. For illustration we plotted the section of the particle trajectory with $R=10$ depicted in Fig.~\ref{fig:R100-1}. \label{fig:Field-lines}}
\end{figure}

We insist on the fact that in this setting the field
lines are integrable and the Poincaré section of the field lines is
depicted in Fig.~\ref{fig:Field-lines}. This section should not
change with the value of $R$ except maybe for the position of resonant
surfaces filled with periodic field lines. We notice that in this
magnetic configuration, the field lines are periodic and purely toroidal
at $r=c$, radius for which the orientation of the winding of field
lines changes.

\subsection{Variation of the magnetic moment $\mu$}

\begin{figure}
\begin{centering}
\includegraphics[width=8cm]{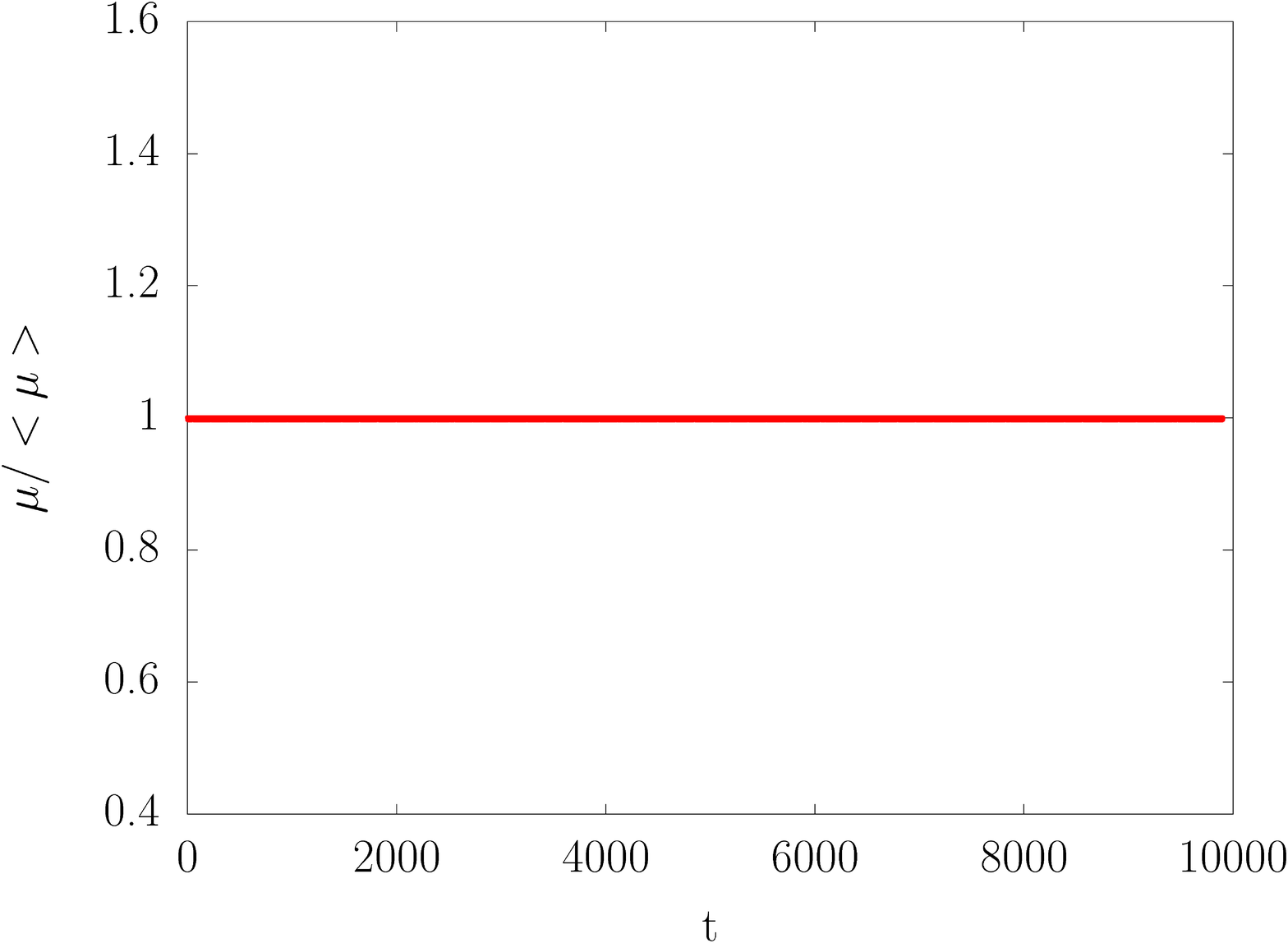}\\
\includegraphics[width=8cm]{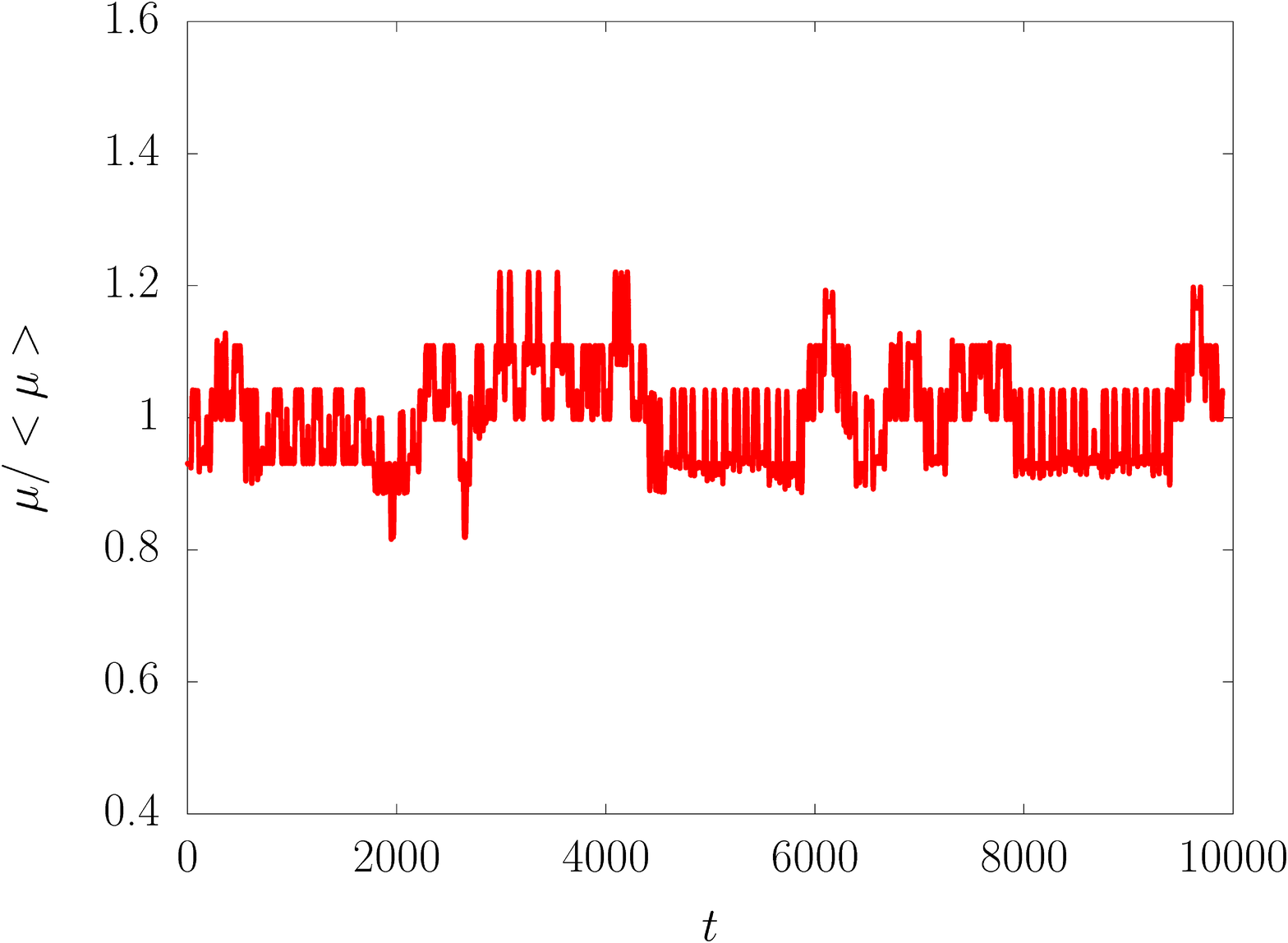}\\
\includegraphics[width=8cm]{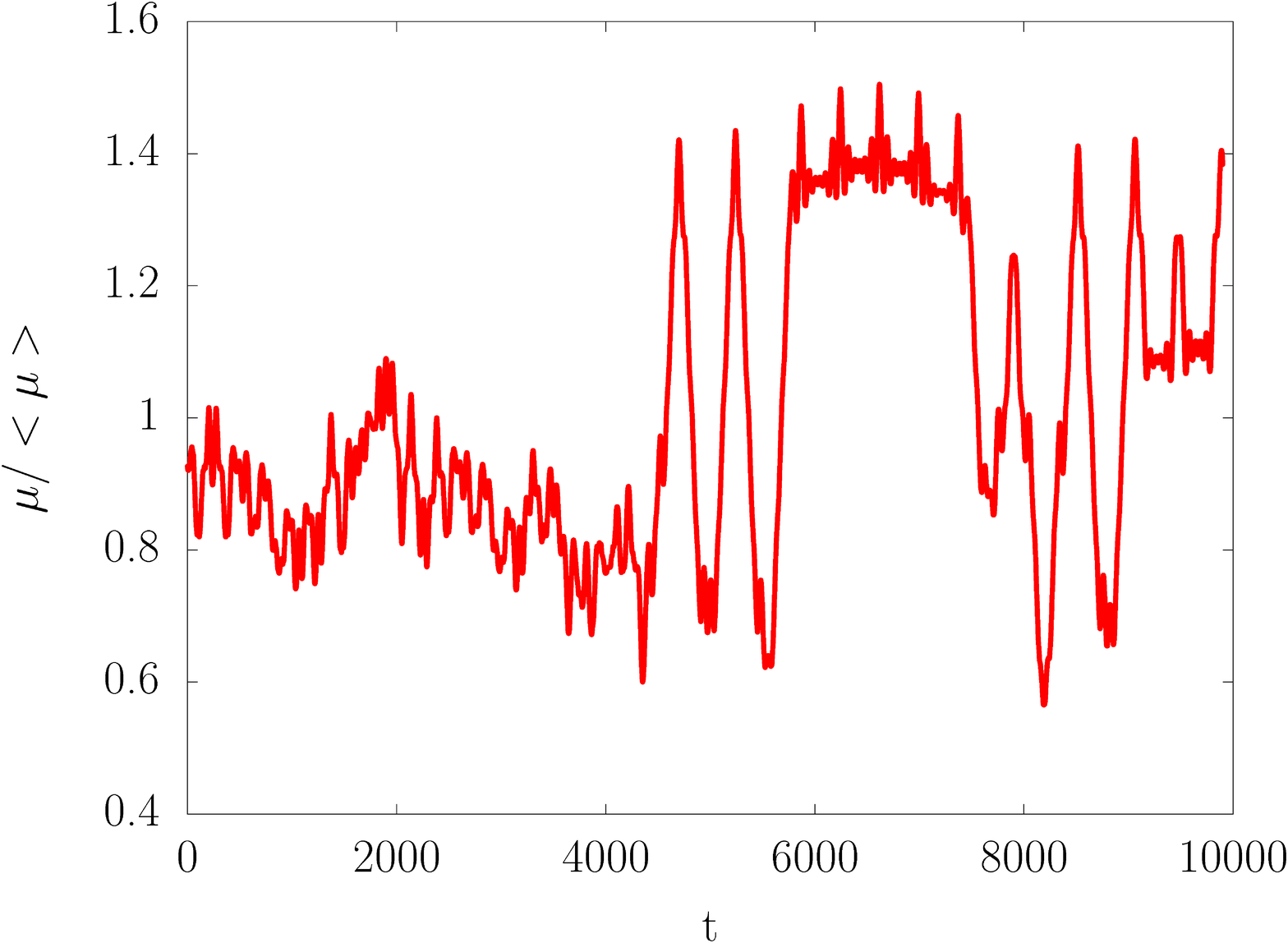}
\par\end{centering}

\caption{Fluctuation of the finite-time averaged magnetic momentum versus time.
The average is performed over about twenty gyration periods. In the
upper plot, the values for regular low energy motion versus time are
represented, $R=10^{4}$, $E=450\: keV$. We can expect that a third
constant of the motion exists. For the middle plot we see fluctuations
of about $\triangle\bar{\mu}/\bar{\mu}\simeq20\%$, we still have
$R=10^{4}$ but higher energy $E=600\: keV$. In the lower plot we
have fluctuations about $60\%$, we are in the case of realistic ratios,
$R=3$ and $E=600\: keV$. In this last case, it is difficult to imagine
that a modified expression of the magnetic moment $\mu$ would become
constant.\label{fig:mu}}
\end{figure}

As previously discussed the presence of chaos in charged
particle trajectory in an axially symmetric system with already two
constant of the motion, implies that the system
is not integrable and as such it precludes the
existence of a third constant of the motion, notably one related to
the magnetic momentum. In order to verify this
statement we measured the variations of the magnetic momentum using
time averages of expression~(\ref{eq:Mu_0}). For this purpose we
compare three different cases in Fig.~\ref{fig:mu}. In this figure,
the average $\bar{\mu}$ are computed on about one hundred cyclotron
gyrations. Except for initial kinetic energy, the
three cases have identical initial conditions. In the first case (upper
plot), we take $R=10000$ and an energy of about $E=450keV$, this
leads to an integrable trajectory, associated to this characteristics,
we note that the variations of $\bar{\mu}$ are small and periodic.
In the second case (middle plot of Fig.~\ref{fig:mu}) we present
the variations of the magnetic moment of a particle when $R=10000$
with an energy $E=600\, keV$, that corresponds to an energy close
to the one of the separatrix and a motion in the chaotic layer. As
seen in Fig.~\ref{fig:R10000_entier}, the motion of this particle
is chaotic and we notice in Fig.~\ref{fig:mu} that the variations
of $\bar{\mu}$ are bigger than in the first case, we actually have
$\triangle\bar{\mu}/\bar{\mu}\simeq20\%$. In the third case, (lower
plot in Fig.~\ref{fig:mu}), we are considering the situation of
the particle moving in a system with the aspect ratios of real tokamaks,
namely $R=3$ at an energy of $E=600keV$. As anticipated by the left
plot in Fig.~\ref{fig:R100-1}, the particle trajectory is  chaotic
and the variations of $\bar{\mu}$ are important $\triangle\bar{\mu}/\bar{\mu}\simeq60\%$
and unpredictable. 
 This seems to confirm that in the phase space of such systems, there
is no global third constant of the motion. The existence of a constant
related to the magnetic moment may be locally true, so we expect the
general picture of a mixed phase space with regions of chaos and regions
with regular motion. This nevertheless poses the problem of a global
reduction performed in for instance gyrokinetics theory in the Maxwell-Vlasov
context.

\section{Conclusion and possible consequences}

Looking for chaos of charged particles trajectories evolving under
the influence of a magnetic field with a toroidal structure, as for
instance an axisymmetric one is not an easy issue. One of the main
difficulties is to visualize trajectories correctly, and the second
problem is to explore the phase space. Given these difficulties, in
this paper we choose to highlight the presence of chaos of particle
trajectory in two specific cases. First, we take into account a ripple
effect, and as such we break a symmetry and the associated exact invariant
of the motion. This allows to insure an a priori non-integrable system
and to localize and visualize chaos more easily. This step allowed
us to insist on the fact that the chaotic motion of charged particles
is not directly linked to the chaotic nature of magnetic field lines,
as indeed we considered a non generic perturbation which kept the
regularity of field lines, while generating chaos in particle trajectories.
So this simple example allowed to underline the fact that the link
between chaos of particle trajectories and chaos of magnetic field
lines is not obvious. In the second case, we considered an axisymmetric
configuration, with two constants of the motion. We managed to observe
chaos by tuning the winding profile of magnetic field lines in order
to create a separatrix in the effective Hamiltonian of the cylindrical
integrable system and perturbing it slowly by adding some curvature
and returning back to the toroidal configuration. We exhibited that
the chaotic region increases when the aspect ratio (major radius)
of the tokamak decreases. At the same time we measured the fluctuations
of the magnetic variations of $\bar{\mu}$ which are also directly
correlated with the chaos of the particle trajectories. The presence
of chaotic trajectories in the case of an axisymmetric field implies
that no third constant of the motion exists as the motion is non-integrable.
This implies therefore that the magnetic moment can not be a global
constant over all phase space, and that this system corresponds likely
to a system with mixed phase phase space, with regions of regular
motion and regions of chaotic motion. We may therefore expect all
the zoology and complex transport properties that is present in one
and a half degrees of freedom system, such as stickiness and associated
memory effects. Another consequence is that one has to be careful
when performing gyrokinetic reduction. Indeed usually the support
of the particle density function described by the Vlasov-Maxwell system
covers the whole phase space. Since there are regions with chaos an
non-constant magnetic moment, a straightforward reduction is a priori
not possible. A future line of research in this field, could be on
how to perform the reduction only in specific regions of phase space,
and how what remains in the other regions is affecting the dynamics
of the reduced distribution. In real case scenarii, it is also likely
that regions where the reduction a priori works may be changing with
the magnetic and electric fields, leading to possible intermittent
reductions, a more thorough study in localizing all these regions
in phase space in realistic tokamak like configurations appears therefore
as a necessity.

\section*{ANNEXE}

In these annexes we used Newton's law, but Noether theorem and a Lagrangian
formalism can be more elegant.

\subsection{Particle trajectory without plasma\label{Annex1}}

In the simple case in which there is no plasma inside tokamak, we
can write the magnetic field as : 

\begin{equation}
\mathbf{B}=\frac{B_{0}R}{\xi}\hat{\mathbf{e}}_{\varphi}\:.\label{eq:Magnetic_solenoid}
\end{equation}
We apply Newton second law to a charged particle of mass $m$, charge
$q$, moving in this magnetic field. In polar coordinates $(r,\theta,z),$
we have :

\begin{gather}
\ddot{r}-r\dot{\theta}^{2}=-\frac{B_{0}Rq}{mr}\dot{z}\:,\label{eq:proj_r}\\
r\ddot{\theta}+2\dot{r}\dot{\theta}=0\:,\label{eq:proj_theta}\\
\ddot{z}=\frac{B_{0}Rq}{mr}\dot{r}\:,\label{eq:proj_z}
\end{gather}
where if we refer to Fig.~\ref{fig:Toric-geometry}, we actually
have used the notations $r=\xi$ and $\theta=\varphi$. We integrate
Eq.~(\ref{eq:proj_theta}) and (\ref{eq:proj_z}) and obtain :

\begin{gather}
r^{2}\dot{\theta}=C\:,\label{eq:const_aire}\\
\dot{z}=\frac{B_{0}Rq}{m}ln(r)+C'\:,\label{eq:vz}
\end{gather}
where $C$ and $C'$ are two constants.

We can then use expression (\ref{eq:const_aire}) and (\ref{eq:vz})
in Eq.~(\ref{eq:proj_r}), multiply it by $\dot{r}$ and integrate
to obtain the effective Hamiltonian

\begin{equation}
\frac{\dot{r}^{2}}{2}+\frac{C^{2}}{2r^{2}}+\frac{(\frac{B_{0}Rq}{m}ln(r))^{2}}{2}+\frac{B_{0}Rq}{m}C'ln(r)=H_{eff_{0}}\:,\label{eq:Heff}
\end{equation}

We end up with an effective Hamiltonian (\ref{eq:Heff}) with one
degree of freedom, and this shows that particle trajectory in this
simple magnetic field is integrable.

\subsection{Particle trajectory in cylindrical geometry\label{Annex2}}

A very similar calculation can be made in the case of a cylindrical
geometry. We recall that the magnetic field is given by 
\begin{equation}
\mathbf{B}=B_{0}\hat{\mathbf{e}}_{z}+f(r)\hat{\mathbf{e}}_{\theta}\:.\label{eq:B_cyl2}
\end{equation}
We apply Newton second law to a charged particle of mass $m$, charge
$q$, moving in this magnetic field. In polar coordinates $(r,\theta,z),$
we obtain:

\begin{gather}
\ddot{r}-r\dot{\theta}^{2}=\frac{q}{m}(B_{0}r\dot{\theta}-f(r)\dot{z})\:,\label{eq:proj_r-1}\\
r\ddot{\theta}+2\dot{r}\dot{\theta}=-\frac{qB_{0}}{m}\dot{r}\:,\label{eq:proj_theta-1}\\
\ddot{z}=\frac{q}{m}\dot{r}f(r)\:,\label{eq:proj_z-1}
\end{gather}
Note that the $z-$direction corresponds to axes along the constant
$\hat{\mathbf{e}}_{\varphi}$ ($R=\infty$), and not the $z$ depicted
in Fig.~\ref{fig:Toric-geometry}. We integrate Eq.~(\ref{eq:proj_z-1}),
and (\ref{eq:proj_theta-1}) as well multiplying it by $r$ before
hand and obtain:

\begin{gather}
r^{2}\dot{\theta}+\frac{eB_{0}}{2m}r^{2}=A\:,\label{eq:const_aire-1}\\
\dot{z}=\frac{e}{m}F(r)\:,\label{eq:vz-1}
\end{gather}
where $A$ is a constant and $F(r)=\int^{r}f(x)dx$.

Now, using expressions (\ref{eq:const_aire-1}) and (\ref{eq:vz-1})
in Eq.~(\ref{eq:proj_r-1}), which we multiply by $\dot{r}$ before
integration, we end up with:
\begin{equation}
m\frac{\dot{r}^{2}}{2}+\frac{mA^{2}}{2r^{2}}+\frac{(qB_{0})^{2}}{8m}r^{2}+\frac{q^{2}}{m}F^{2}(r)=H_{eff}\:,\label{eq:Heff-1}
\end{equation}
in which $H_{eff}$ is a constant. This system is also integrable,
but contrary to the Hamiltonian~(\ref{eq:Heff}) presents the interest
to allow us to create a separatrix by tuning the function $F(r)$,
in other words the so called $q-$profile.
\begin{acknowledgments}
We are very grateful to P. J. Morrison for providing useful comments
and remarks. This work was carried out within the framework the European
Fusion Development Agreement and the French Research Federation for
Fusion Studies. It is supported by the European Communities under
the contract of Association between Euratom and CEA. The views and
opinions expressed herein do not necessarily reflect those of the
European Commission.
\end{acknowledgments}
\bibliographystyle{apsrev4-1}
\
\end{document}